# A deep learning-enabled smart garment for versatile and accurate sleep conditions monitoring in daily life

## Authors

Chenyu Tang[†,1], Wentian Yi[†,1], Muzi Xu[1], Yuxuan Jin[2], Zibo Zhang[1], Xuhang Chen[4], Caizhi Liao[1], Peter Smielewski[3], Luigi G. Occhipinti[*,1]

## Affiliations

[1]Electrical Engineering Division, Department of Engineering, University of Cambridge, UK
[2]Cavendish Laboratory, University of Cambridge, UK
[3]Department of Clinical Neurosciences, University of Cambridge, UK

[†]These authors contributed equally: Chenyu Tang, Wentian Yi
[*]Correspondence to: Luigi G. Occhipinti, email: lgo23@cam.ac.uk

**Abstract:** In wearable smart systems, continuous monitoring and accurate classification of different sleep-related conditions are critical for enhancing sleep quality and preventing sleep-related chronic conditions. However, the requirements for device-skin coupling quality in electrophysiological sleep monitoring systems hinder the comfort and reliability of night wearing. Here, we report a washable, skin-compatible smart garment sleep monitoring system that captures local skin strain signals under weak device-skin coupling conditions without positioning or skin preparation requirements. A printed textile-based strain sensor array responds to strain from 0.1% to 10% with a gauge factor as high as 100 and shows independence to extrinsic motion artefacts via strain-isolating printed pattern design. Through reversible starching treatment, ink penetration depth during direct printing on garments is controlled to achieve batch-to-batch performance variation < 10%. Coupled with deep learning, explainable artificial intelligence (XAI), and transfer learning data processing, the smart garment is capable of classifying six sleep states with an accuracy of 98.6%, maintaining excellent explainability (classification with low bias) and generalization (95% accuracy on new users with few-shot learning less than 15 samples per class) in practical applications, paving the way for next-generation daily sleep healthcare management.

# I. Introduction

Sleep is a vital component of human health, occupying about one-third of daily life. However, over 60% of adults experience poor sleep quality, leading to significant personal and economic consequences, including the loss of 44 to 54 working days annually and a global GDP reduction estimated between 0.64% and 1.31% [1, 2, 3]. Sub-healthy and high-risk sleep patterns — such as mouth breathing, snoring, bruxism, and sleep apnea — are major contributors to poor sleep quality [4, 5, 6, 7] and are linked to chronic diseases like cardiovascular disease, diabetes, and emotional disorders [8, 9, 10, 41]. Thus, effective monitoring and identification of these sleep states are crucial for modern health management.

Traditional sleep monitoring, primarily through polysomnography (PSG), while accurate, is not feasible for home or long-term use due to its complexity and cost [11]. This limitation has spurred the development of portable, user-friendly alternatives. Smart wearable devices have emerged as promising solutions [12, 13, 14], integrating sensors such as photoplethysmography (PPG) within watches or wristbands [15, 16, 17]. However, these devices often lack the capacity to capture comprehensive physiological data needed for a thorough analysis of various sleep states [18]. Innovative designs using physical sensors like humidity, mechanical, and acoustic sensors have attempted to bridge this gap [19, 20, 21, 22, 23]. Although they offer broader physiological insights, they typically require multiple integrated sensors, which increases bulkiness and energy consumption, compromising comfort and long-term usability [19, 21, 22]. Another approach involves integrating electrophysiological sensors, such as electroencephalograms (EEG), electrooculograms (EOG), and electromyograms (EMG), into facial or ear areas [24, 25, 26]. While improving comfort and data richness, these non-invasive electrophysiological sensors still face challenges with inherent artefact noise, limiting accuracy [27, 28]. Thus, there is still a lack of versatile technology with both good comfort and high precision for continuous monitoring of various sleep conditions in real world scenarios.

Here, we developed a smart garment that balances the trade-off between versatility, comfort, and accuracy for monitoring sleep health in daily environments. The smart garment is integrated with an ultrasensitive sensor array directly printed on the collar, capable of identifying multiple sleep patterns. Direct printing on textile substrates enables integration of multi-functional electronic elements directly onto cloths with scalability and design flexibility. The developed smart garment is able to collect mixed-mode signals generated by vibrations from various sleep activities such as breathing, snoring, teeth grinding, and sleep apnea, which are transmitted to the extrinsic laryngeal muscles from multiple anatomical locations including the velum, oropharynx, tongue, and epiglottis (Figure 1a). Utilizing a multi-channel graphene textile strain sensor array, screen printed at its collar based on principles of ordered cracks and selective starching treatments, the smart garment leverages ultrahigh sensitivity (gauge factor >100), scalability (±20% conductivity fluctuation), and durability (stable over 10,000 cycles of stretching tests) to continuously monitor subtle vibrations of the extrinsic laryngeal muscles while ensuring user comfort. Additionally, our designed strain isolation treatment allows to adjust the substrate's rigidity, isolating strain artefacts caused by nocturnal turnings and other motions outside the sensing area, thereby significantly reducing the magnitude and drift of motion artefacts. Additionally, due to its multi-channel design, the smart garment can be easily used by wearers in a positioning-free manner in real-world application scenarios. The designed deep learning model, SleepNet, uses the captured signals to accurately identify six sleep states ranging from healthy to sub-healthy to high-risk, including nasal breath, mouth breath, snoring, bruxism, central sleep apnea (CSA), and obstructive sleep apnea (OSA), achieving an accuracy of 98.6% along with decent inference speed (Figure 1b). Explainable artificial intelligence (XAI) visualizations confirm that the model comprehensively understands the sleep patterns, avoiding biases towards noisy regions, thus demonstrating its robustness. Moreover, transfer learning tests show that after few-shot learning (with only 15 samples per class), the model can achieve up to 95% classification accuracy on new users, showcasing the system's powerful generalization capabilities. In Table 1, we have summarized the functions and features of our system compared with other wearable sleep monitoring systems in the literature, showing that, to the best of our knowledge, the proposed system consistently outperforms state-of-the-art solutions in functionality performance, reliability, versatility and user-convenience. Given its user-friendly design, high accuracy, and personalization features, the proposed technology favors adoption of this smart garment for daily use, empowering individuals to better understand and manage their sleep health and promoting well-being for a wide range of users.

## II. Results

**Printed strain sensor array on garments**

Figure 2a illustrates the multi-layered structure of the strain sensing array screen-printed on a high-neck top made of elastic knitted fabric. In the multi-layer screen printing process, printing quality can be easily adapted to different textile substrates by tuning the printing parameters and ink properties. To ensure stable performance, we introduced two different treatments in the fabrication process: starching treatment using sodium carboxymethyl cellulose (CMC Na), for the sensing area, and strain isolation treatment using polyurethane acrylate (PUA), for the surrounding regions. Both cellulose derivatives and acrylates are common starching agents in garment printing industry [29]. CMC Na is a water-soluble polymer derived from cellulose known for its film-forming ability, which provides a smooth and uniform surface [30]. CMC Na also improves the adhesion of the printed graphene ink to the substrate, reducing the likelihood of delamination under mechanical stress. The PUA provides high stiffness and excellent adhesion by forming a robust, cross-linked network upon UV exposure [31]. Introducing a rigid PUA layer in the textile strain sensor array modifies the rigidity of selected area on the textile, redistributing strain caused by body movements during sleep (Supplementary Figure 1). The isolated area remains inert to large-scale uniaxial stretching, ensuring that local strain is accurately measured without interference, as shown in Supplementary Figure 3. The UV-curable nature also prevents clogging of the screen during the printing process, extending the lifespan of the printing screen and maintaining consistent printing quality.

After the starching and strain isolation treatment, the crossbar silver electrodes, separated by an insulating layer, are printed, followed by the graphene sensing layer on the surface, which comprises exfoliated graphene flakes bound with ethyl cellulose (EC). The graphene flakes production process is tuned to obtain flakes under 1 μm lateral dimension to ensure stable dispersion in the graphene ink formulation and avoid mesh clogging (Supplementary Figure 4). Controlling the formation of ordered cracks in the graphene layer is critical for fabricating crack-based strain sensors with repeatable performance [32]. During the screen-printing process, the ink is squeezed through the mesh by the squeegee, resulting in a patterned thin film onto the textile substrate. The stress concentration at the boundaries of the textile structural units induces the formation of regular cracks (Figure 2b). This process does not require complex pre-stretching or pre-treatment steps, making it compatible with conventional printing processes adopted by the garment manufacturing industry [42]. A known issue in textile printing processes is related to capillary forces causing the ink to spread, while air pockets block ink deposition, leading to variability and poor printing quality. Additionally, in this case, excessive penetration of graphene ink into the textile can create a graphene/textile composite which is insensitive to strain and acts as an extra conductive pathway to the surface cracking layer (Supplementary Figure 5). To overcome these limitations and inhibit ink penetration and air trapping during the printing process, another starching treatment with CMC Na was used, which creates a controlled surface for ink deposition, preventing deep penetration and ensuring the graphene forms a brittle surface layer that cracks under strain. As shown in Supplementary Figure 7, the penetration depth of graphene ink in starched textile is significantly lower compared to untreated textile. The fabrication process is illustrated in Supplementary Figure 8. It is scalable and compatible with industrial textile printing processes, making it suitable for mass production of smart garments [43].

The graphene/EC coating with ordered cracks shows a reliable linear response to small uniaxial strains from 0.1% to 5% with a gauge factor over 100 (Figure 2c). Moreover, the strain sensors demonstrate a rapid response to straining cycles with frequency ranging from 1Hz to 10 Hz (Figure 2d), enabling the real-time monitoring of fast and subtle vibrations produced by the throat during sleep. The durability of the graphene sensor was tested through tensile tests at 1% uniaxial strain and 1 Hz frequency, showing consistent performance (Figure 2e). In contrast to the high response of the strain-sensing layer, the stretchable silver electrodes exhibit high conductivity with negligible strain response, ensuring stable electrical connections (Figure 2f). To ensure consistency and reliability, we studied the performance distribution of two-terminal resistance and gauge factor across 50 strain sensor units. A resistance variation of less than 12.7% and a gauge factor variation of less than 9.2% were achieved by controlling printing conditions and applying starching treatments, as shown in Figure 2g. Moreover, as observed from Figure 2i, compared to the device without PUA treatment, the device with treatment shows significantly mitigation on the amplitude and drift of artefact signal during nocturnal turnings, which validates the effectiveness of strain isolation treatment.

The washability of the sensors was tested by immersing the devices in water under 500 rpm stirring for 5 to 60 minutes, drying at room temperature [44]. The performance degradation after washing was less than 10%. Similar results were observed when detergent was added (Figure 2h). Additionally, the smart garment substrate maintains excellent breathability, even after treatment and integration of the sensing devices, with moisture vapor transmission rate (MVTR) values over 10 times higher than Tegaderm [45], a commonly used medical dressing substrate produced by 3M (Supplementary Figure 9). Furthermore, the strain sensor array results in no skin irritation or other side effects after 8 hours of wear, as the typical duration of an overnight sleep (Supplementary

Figure 10). Artificial sweat was used to test the resistance of the smart garment against prolonged use in sweat environment (ISO 3160-2 standard), resulting in no signs of degradation of its electrical resistance even after 2 hours exposure to both normal (1 hour at 0.32 mg/cm$^2$/min) and high (1hour at 2.7 mg/cm$^2$/min) sweat rate (Supplementary Figure 11).

**Positioning-free monitoring of sleep conditions with the array**

We employed a six-channel strain sensor array integrated into the collars of textile garments, positioned around the participants' necks. This setup was designed to collect minute vibrational signals from the extrinsic laryngeal muscles, which vary according to different sleep conditions (Figure 3a). Six channels provide sufficient spatial resolution to capture detailed vibrational patterns associated with various sleep conditions. A larger number of sensors would increase complexity and data processing demands with no significant gain in accuracy, while a smaller number might fail to capture critical signals. This configuration allows for reliable detection of sleep states across a diverse range of users. Utilizing a multiplexer, we could read the responses from a circular six-channel piezoresistive strain sensor arranged in a crossbar structure for further analysis. In our design, each circular sensing channel can be equivalently viewed as a parallel connection of four quarter-ring variable resistors, as shown in Figure 3b, when we reflect the resistance values of the vertical and horizontal lines of the sensors. This design provides two-dimensional sensitivity to both horizontal and vertical strain. The sensor area was optimized by arranging the channels to maximize coverage around the neck, ensuring that at least one sensor captures the strongest response, regardless of garment positioning. This layout minimizes the need for precise positioning and enhances usability, allowing users to wear the garment comfortably without worrying about keeping the sensors aligned with the breathing-related sensitive areas of the neck region. Figure 3c illustrates the response signals during a 10-second sample of nasal breathing, captured by the six channels at standard wearing positions. Correlation analysis revealed that although the intensity of the strain responses varied across different channel locations covering the throat area, the signal characteristics were highly correlated (Pearson correlation coefficients greater than 0.9 between any two channels). This pattern persisted even when the sensor array was worn askew, not in the standard position (see Supplementary Figure 12). These findings demonstrate that our design, which utilizes a relatively large coverage area of the strain sensor array, essentially ensures that the region with the strongest response falls within the device's sensitive area. Consequently, our sleep conditions monitoring system does not require precise positioning, implying high resilience to positional variances, which ensures the practicality of this monitoring system in real-world application scenarios. Furthermore, due to the high correlation between channels, subsequent neural network-based pattern recognition needs only to consider the channel with the strongest response as representative for signal processing and extraction of the relevant features, which improves the network's inference efficiency while ensuring accuracy.

To ensure effectiveness of the proposed approach, we built a comprehensive dataset encompassing six distinct classes of sleep patterns: nasal breathing (normal healthy condition), mouth breathing, snoring, bruxism, CSA, and OSA. The dataset was collected from seven healthy subjects and spans a health spectrum from healthy and sub-healthy to high-risk categories. Therefore, for the three conditions—bruxism, OSA, and CSA—which are rare among healthy individuals, the data were augmented through simulated conditions under the guidance of medical experts. In contrast, data for mouth breathing, nasal breathing, and snoring were collected from actual sleep sessions in different subjects. Detailed protocols for data collection are described in the Methods section. As visualized in Figure 3d, the temporal and spectral characteristics of these sleep conditions were meticulously analyzed focusing on the signals emanating from the channel exhibiting the strongest response. We observed that the effective vibrational signals originating from the extrinsic laryngeal muscles are predominantly found within the low-frequency domain, specifically below 10 Hz. This frequency band captures the physiological nuances of each sleep pattern, enabling a precise delineation of the conditions. Nasal and mouth breathing exhibit fundamental differences in their spectral signatures, reflecting variations in airflow mechanics and potential diagnostic markers for respiratory efficiency. The irregular and prominent vibrational patterns associated with snoring suggest the presence of disrupted airflow dynamics and serve as indicators of upper airway resistance. Notably, the episodic high-amplitude signals of bruxism provide clear evidence of nocturnal teeth grinding, which could be associated with high stress levels or sleep disturbances. The absence of vibrational activity during respiratory pauses in CSA and the erratic signal fluctuations indicative of breathing efforts against obstruction in OSA are consistent with the clinical understanding of these two conditions. More visualization of the samples is shown in Supplementary Figure 13-18.

**Sleep conditions recognition with a deep learning model**

The preprocessing involves labeling (see detailed protocols in Supplementary Note 2), segmentation (10s signals were segmented into one sample, and 50% overlapping was introduced for data augmentation), and Z-score normalization [33]. The 10-second interval for segmentation was chosen to balance capturing essential temporal features of sleep events with computational efficiency. This duration effectively captures the necessary contextual information for identifying patterns like breathing cycles and snoring while avoiding the noise and inefficiency of longer segments, thus ensuring accurate and real-time processing. To enhance model robustness and data augmentation, a 50% overlapping across adjacent time series was used, which increases the number of training samples and improves the model's ability to generalize across varied sleep patterns. Z-score normalization was applied to standardize the data, eliminating biases caused by amplitude variations and ensuring consistent input scales for accurate pattern recognition.

After that, signals from the channel with the strongest response were fed into our specially designed deep learning model (SleepNet) for sleep conditions recognition (Figure 4a). The SleepNet is composed of three core components: First, the learnable positional encoder adopts a residual bidirectional LSTM (BiLSTM) framework to understand the sequential nature of the input data. This approach surpasses traditional sinusoidal positional encoding by dynamically learning positional information, which is particularly advantageous in sleep pattern analysis where the temporal relationship between events can signify different breath cycles or disturbances [34]. Second, the multi-head self-attention module, derived from Transformer architectures, enables nuanced discrimination of significance across the data sequence. By effectively utilizing long-range dependencies, the model ensures a comprehensive understanding of the sequential data, reflecting the true complexity of sleep behaviors over time [35]. Lastly, the one-dimensional Residual Network (ResNet) layers function as a hierarchical feature extractor. With their ability to skip connections, they prevent the vanishing gradient problem and enhance the flow of information, thus allowing the model to learn more complex patterns effectively. The 1D convolution within these layers ensures that the model is tailored to handle time-series data, offering a nuanced analysis of temporal patterns during sleep [36].

Figure 4b displays the confusion matrix for the model's classification of sleep patterns. The model achieved impressive classification results: the accuracy for each category was above 95%, with an overall accuracy of 98.6%. In the comparative experiments shown in Figure 4c, SleepNet surpassed the performance of its individual component architectures, such as the Transformer and 1D ResNet, as well as other state-of-the-art models in terms of accuracy. Additionally, it's noteworthy that, after pruning the least important 50% of the nodes in the 1D-ResNet module of the model and re-training to obtain a pruned SleepNet, not only did the number of floating-point operations per second (FLOPS) –a critical indicator of model inference speed and energy efficiency– decrease by 30%, but there was also a slight increase in accuracy. This can be attributed to the phenomenon known as "pruning-induced efficiency", where removing redundant or less important nodes can lead to a more streamlined and efficient network. The re-training phase helps the model to re-allocate its resources towards the most salient features, potentially improving generalization and thus accuracy [37]. Overall, the pruned SleepNet achieved less than 0.5 GFLOPs and fewer than 5 million parameters, highlighting its capability for deployment on edge computing devices. This efficiency allows the model to run effectively on low-power, resource-constrained hardware, making it suitable for real-time sleep monitoring in wearable devices without compromising performance.

Figure 4d illustrates the process of hyperparameter optimization for the model. It can be observed that in the majority of hyperparameter combinations, the model exhibits high accuracy (greater than 90%). This indicates that the model's performance is not overly sensitive to the specific values of its hyperparameters, demonstrating the model's robustness. Moreover, such robustness might also imply that the fundamental features learned by the model are strong predictors of sleep patterns, allowing for decent performance despite variations in model configuration. Figure 4e shows the ROC curves for the model's classification of each sleep pattern type. The AUC values are almost equal to 1 for each classification task, indicating that the model is effective and has achieved satisfactory classification performance.

SmoothGrad visualizations in Figure 5a provide a clear depiction of how different segments of the signal contribute to the model's classification decisions [38]. The shading intensity represents the degree to which each point in the time series influences the output, with darker shades indicating higher importance. For instance, the "Nasal Breath" and "Mouth Breath" classifications exhibit consistent contribution patterns throughout the breathing cycle, reflecting the model's reliance on rhythmic features for these classes. Conversely, "Snoring" shows a more variable contribution pattern, which likely corresponds to the erratic nature of snoring events. The "Bruxism" class demonstrates pronounced contributions at peaks, which may correspond to teeth grinding instances. In the cases of "CSA" and "OSA", the model identifies critical contributions at pause cycle and obstruction cycle. This distribution of contributions is consistent with established physiological patterns, which means that the model gives appropriate weight to relevant features across different classes, reflecting a balanced understanding of the data rather than an overreliance on certain input aspects that could lead to skewed predictions (e.g. noise). Furthermore, this underscores the model's capacity to not only recognize but also assign appropriate significance to the distinct temporal features within the complex landscape of sleep-related signals, enhancing the interpretability of its predictions. Figure 5b compares the distribution of raw data with the features extracted by the model in the t-SNE plane, illustrating the model's ability to discern and delineate the complex structure within the data. The t-SNE visualization of extracted features shows distinct clusters corresponding to different sleep patterns, which are not easy discernible in the raw data. This indicates the effectiveness of the feature extraction in capturing the patterns and underlying relationships necessary for accurate classification.

To assess the model's generalization ability, we conducted transfer learning tests, applying the model trained on five participants to the dataset of two new participants. As shown in Figure 5c, with only 15 samples per category for few-shot learning, the model achieved an accuracy of up to 95% on the new subject. In contrast, training only on the new participant's dataset without transfer yielded a few-shot learning accuracy of merely 80%. This reflects the model's ability to adapt and maintain performance across different individuals, showcasing its robust transferability and capacity to leverage previously learned patterns to quickly adapt to new, unseen data.

## III. Discussion

Decoding human sleep patterns is important yet complex. Despite the promising development of wearable devices for sleep health monitoring in recent years, creating a system that combines versatility, comfort, and accuracy remains a significant challenge, hindering adoption. In this work, we have designed a smart garment integrated with a six-channel strain sensor array. This ultrasensitive strain sensor array, characterized by excellent robustness and durability, can collect subtle vibrations from the extrinsic laryngeal muscles associated with various sleep patterns, and its multi-channel design eliminates the need for positioning due to its spatial resolution. Specifically, the strain isolation treatment mitigates strain artefacts caused by nocturnal movements and other motions outside the sensing area, which ensures the long-term reliability and robustness of the smart garment during overnight use. Despite utilizing only a single modality of strain response signals, our smart garment, equipped with a customized deep learning neural network, can comprehensively analyze and recognize subtle vibrations originating from various physiological sites and transmitted to the extrinsic laryngeal muscles. It accurately classifies six sleep patterns: nasal breath, mouth breath, snoring, bruxism, CSA, and OSA. Additionally, it can efficiently and effectively adapt to new users, maintaining high accuracy in its classifications. We believe that our smart garment offers a promising solution for versatile sleep monitoring in wearable devices, suitable for broad consumer electronics market to provide ongoing sleep monitoring for general users.

An important direction for future work involves conducting broader user studies with the smart garment across diverse populations. While our study demonstrated the system's effectiveness among healthy users of varying body types and gender within a young demographic, it would be valuable to investigate whether the micro-vibration patterns of the extrinsic laryngeal muscles during different sleep states exhibit consistent characteristics across a wider range of age groups, including both healthy individuals and patient populations. Such exploration would not only enhance our understanding of the system's applicability but also provide deeper insights into the fundamental principles of sleep. If current strain signals prove insufficient to develop a universal model for more diverse groups, integrating demographic characteristics (e.g., age, gender, body type) and prior information (e.g., key daytime activities) to create personalized models would be a meaningful approach [14]. Additionally, while this study has quantified the feasibility of the system for real-time monitoring in terms of FLOPs and parameter size, future work should focus on integrating the smart garment into a truly wireless edge-computing system addressing low energy consumption and latency requirements for application in real-world settings. With these further developments, the proposed smart garment has the potential to become a new benchmark in sleep monitoring technology.

## IV. Methods

**Materials**
TIMREX KS 25 Graphite (particle size of 25μm) was sourced from IMERYS. Stretchable conductive silver ink was obtained from Dycotec Materials Ltd. Ethyl cellulose and sodium carboxymethyl cellulose were purchased from SIGMA-ALDRICH. Flexible UV Resin Clear was acquired from Photocentric Ltd. The textile substrate, composed of 95% Polyester and 5% spandex, was procured from Jelly Fabrics Ltd.

**Ink Formulation**
The graphene ink for screen printing was prepared following a reported method. Briefly, 100g of graphite powder and 2g of ethyl cellulose (EC) were mixed in 1L of isopropyl alcohol (IPA) and stirred at 3000 rpm for 30 minutes. The mixture was then added into a high-pressure homogenizer (PSI-40) at 2000 bar pressure for 50 cycles to obtain graphene dispersion. The graphene dispersion is centrifuged at 5000g for 30 min to remove unexfoliated graphite. To prepare the CMC Na starching solution, CMC Na was dissolved in water at 20% wt. concentration.

**Fabrication of Textile Strain Sensor Arrays**
The textile substrate was washed with detergent, thoroughly dried, and then treated with UV-ozone for 5 minutes to clean the surface. Screen printing was performed using a 165T polyester silk screen on a semi-automatic printer (Kippax & Sons Ltd.) set with a squeegee angle of 45 degrees, a spacer of 2mm, a coating speed of 10mm/s, and a printing speed of 40mm/s. Printing pressure was pneumatically controlled, with higher pressure applied for the viscous starching agent, and moderate pressure for the thinner graphene ink silver ink to reduce penetration. After each printing pass, the textile was blown to dry. After printing, the sensor was washed with water to remove CMC Na and dried at 80 °C overnight. A biaxial strain of around 10% was then applied to induce the formation of ordered cracks.

**Characterization of Structure and Performance**
The size distribution of graphene flakes was analyzed using a Bruker Icon Atomic Force Microscope (AFM) in an area approximately 20 μm × 20 μm. Scanning Electron Microscopy (SEM) images were taken with a Magellan 400, after sputtering the textile samples with a 5 nm layer of gold to enhance conductivity. Optical images were captured using an Olympus microscope.

**Tensile Tests**

Tensile properties of the textile strain sensors were evaluated using a Deben Microtest 200N Tensile Stage and an INSTRON universal testing system. Electrical signals were recorded concurrently with a potentiostat (EmStat4X, PalmSens) and a multiplexer (MUX, PalmSens). Copper tape was crimped onto the contact pads of the samples, supplemented with a small amount of silver paste to improve electrical contact.

**Experimental setup of data acquisition**

Our strain sensors were screen printed onto the collars of garments, equipped with copper strips at the crossbar electrodes. For data acquisition, we employed a potentiostat (EmStat4X, PalmSens) and a multiplexer (MUX8-R2, PalmSens) as our primary readout modules. These modules consistently supplied a 1V voltage, with the resulting output being the current passing through the strain sensors. We selected a sampling frequency of 100Hz and segmented the data into 10-second samples for detailed analysis. Our data collection process was specifically crafted to reflect real-world conditions, accommodating variations in the positioning and tightness of the collar with each use. Throughout our extensive data collection from various participants, we intentionally avoided strict calibration of the collar's position or tightness. Participants were advised to wear the smart garment comfortably and put the collar around their necks, ideally positioned at the mid to upper throat level. This method ensured that our dataset represented a wide range of real-life scenarios, capturing the inherent variability in the collar positioning and tightness across different users and experimental setups.

**Sleep conditions dataset collection**

Since all our participants were healthy students recruited from the University of Cambridge (7 students, average age 25, 4 males and 3 females), complying with the University of Cambridge Engineering Department's Ethical Approval for the Research Project: "Wearable Sensor System for Breath Monitoring", our subjects did not exhibit significant symptoms of Bruxism, CSA, or OSA. Therefore, the three types of sleep conditions were simulated following training under the guidance of medical experts. For the collection of bruxism, we included simulated instances of grinding, clenching, and tapping; for CSA, we instructed participants in voluntary end-expiratory central apnea during breathing; for the more challenging simulation of OSA, we trained participants to utilise Muller maneuver to maintain a lower intrathoracic pressure [39] and followed the characteristic descriptions of reports from the American Association of Sleep Medicine (AASM), introducing clinical $SpO_2$ as an auxiliary simulation tool, marking a segment as a valid OSA pattern only when $SpO_2$ continuously fell below 90% within the sample, or a continuous decrease of more than 4% from baseline [40]. These simulations were carefully developed with clinicians to ensure the primary mechanics of airway obstruction and breath maneuver conform realistic situations. The data for the other three conditions (nasal breath, mouth breath, and snoring) were collected during actual sleep states. To maintain consistency with the daily environments, sleep positions are freely chosen by the subjects. In total, we collected 2119 samples, including 728 samples of nasal breath, 701 samples of mouth breath, 262 samples of snoring, 180 samples of bruxism, 102 samples of CSA, and 146 samples of OSA. See more detailed data collection and annotation protocol in Supplementary Note 2. The participants had varying body shapes, adding diversity to the study (see details in Supplementary Table 1).

**Software environment**

Signal preprocessing was performed on a MacBook Pro equipped with an M1 Max CPU. Network training was conducted using Python 3.8.13, Miniconda 3, and PyTorch 2.0.1 in a performance-optimized environment. Training acceleration was enabled by CUDA on NVIDIA 4090 GPU.

**Data availability**

The datasets supporting this study will be available from the GitHub repository before publication.

**Code availability**

The code supporting this study will be available from the GitHub repository before publication.

**Acknowledgments**

C.T. was supported by Endoenergy Systems (grant No. G119004) and Haleon (grant No. G110480), W.Y. was supported by Pragmatic Semiconductor (grant No. G117793) and Haleon (grant No. G110480), L.G.O. acknowledges funding from EPSRC (grants No. EP/K03099X/1, EP/L016087/1, EP/W024284/1, EP/P027628/1), the EU Graphene Flagship Core 3 (grant No. 881603), and Haleon (grant No. G110480). We would like to thank Dr. Ian Smith from Royal Papworth Hospital, UK, and Steve Walsh from Haleon for their valuable advice on the project.

**Table 1** | Comparison of the proposed smart garment's features with state-of-the-art wearable sleep monitoring systems.

|  | **This Work** | Shen *et al.* [16] | Yue *et al.* [19] | Sun *et al.* [21] | Tarim *et al.* [22] | O'Hare *et al.* [23] | Kwon *et al.* [24] |
|---|---|---|---|---|---|---|---|
| Form factor | Smart textile garment | Smart bracelet | Philtrum and neck e-skin | Philtrum and chest patch | Wearable platform | Masseter EMG patch | Smart face patch |
| Sensors | 6-channel strain sensor | 1 PPG sensor | 1 pressure, 1 humidity, 1 temperature sensor | 1 pressure and 1 humidity sensor | 1 IMU and 1 temperature sensor | 1 bipolar surface EMG | 2-channel EEG, 2-channel EOG, 1-channel EMG |
| Functions | Mouth breath detection | Sleep apnea detection | Sleep stage classification | Sleep apnea detection | Breath patterns detection | Bruxism detection | Sleep apnea detection |
|  | Snoring detection |  |  |  |  |  | Sleep quality assessment |
|  | Bruxism detection |  |  |  |  |  |  |
|  | Sleep apnea detection |  |  |  |  |  |  |
| Scalability | Good | Good | Bad | Bad | Good | Good | Good |
| Robustness | Good | Medium | Bad | Bad | Medium | Medium | Medium |
| Durability | Good | Good | Good | Good | Good | Bad | Good |
| Washability | √ | × | × | × | × | × | √ |
| Breathability | Good | Good | Medium | Bad | Good | Bad | Medium |
| No precise positioning | √ | √ | × | × | × | × | × |
| Overall accuracy | 98.6% | 81.82% | N/A | N/A | 96% | 82.8% | 88.5% |
| Generalization ability | Good | Good | Good | Good | Good | Medium | Medium |
| Explainability | Good | Good | Good | Good | Good | Bad | Bad |

**Note:** N/A indicates that this feature was not tested in the literature and does not necessarily imply that the method lacks this capability in principle.

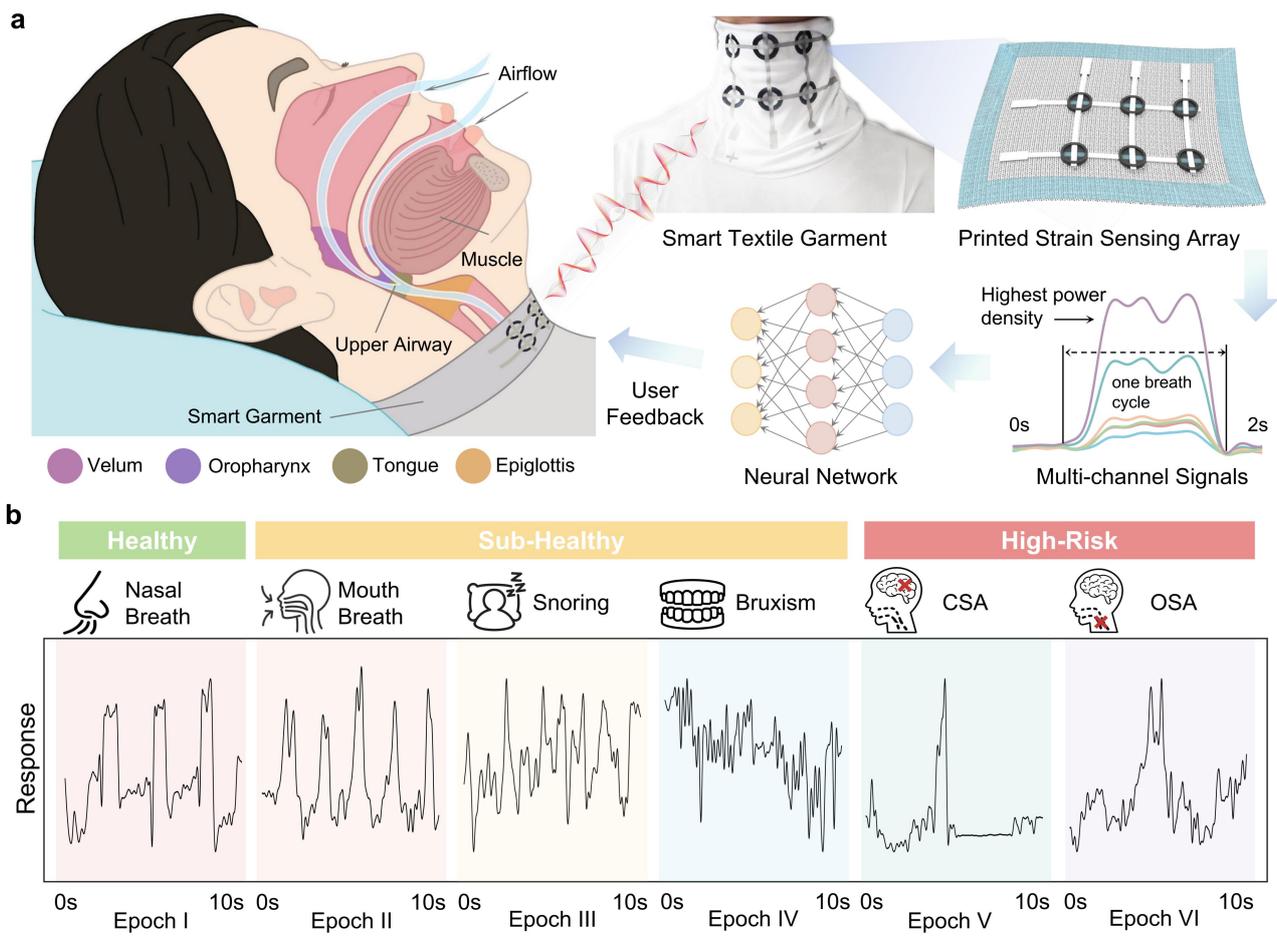

**Figure 1 | Overview of the smart garment system for versatile sleep behavior monitoring. a.** The monitoring of sleep behavior begins by detecting subtle vibrations at the extrinsic laryngeal muscle, which are induced by physiological vibrations emanating from various anatomical locations such as the velum, oropharynx, tongue, and epiglottis. These vibrations are then captured by a six-channel strain sensor array printed onto the collar of a garment. The signals from the channel with the strongest response are processed by a deep learning neural network, SleepNet, which is designed for recognizing and analyzing sleep patterns. **b.** Visualization of the signals of six different sleep patterns (channel with the highest power density).

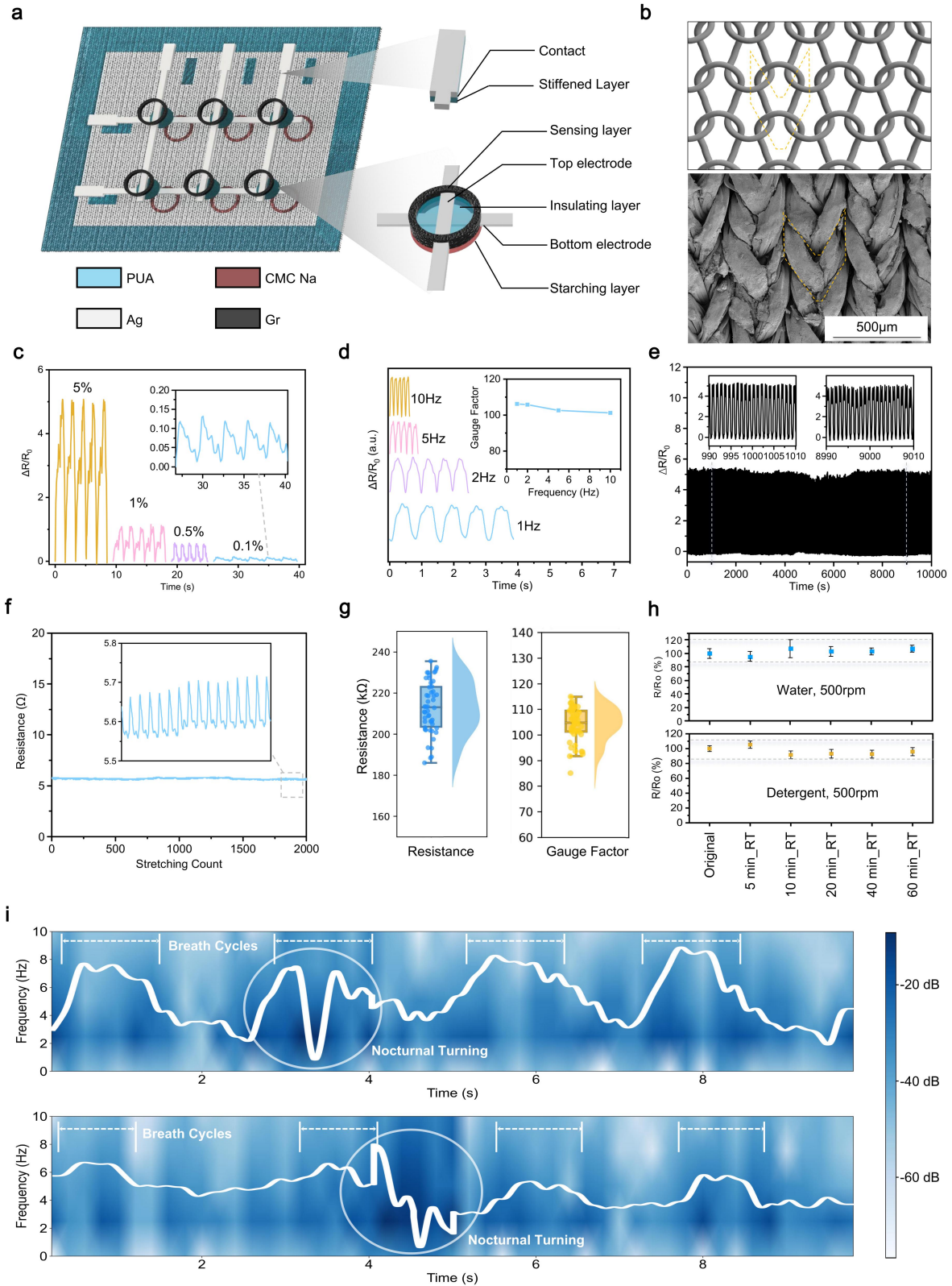

**Figure 2 | Characterization of the device. a.** Schematic of the strain sensor array, including the elastic spandex textile substrate, starching layer, crossbar electrodes, and circular sensing layer. **b.** Schematic (top) and SEM image (bottom) of the ordered cracks formed around the textile structural units after graphene printing. The structural units are labeled with yellow dashed lines. **c.** Resistance response to cyclic tensile strains of 5%, 1%,

0.5%, and 0.1%. Inset shows a zoomed-in view of the response at 0.1%. **d.** Dynamic response test under uniaxial cyclic tensile at 1% strain with different frequencies: 10 Hz, 5 Hz, 2 Hz, and 1 Hz. Inset shows the gauge factor at each frequency. **e.** Durability test of graphene strain sensors under 10,000 cycles of 1% strain. **f.** Strain response of stretchable silver electrodes under 1% strain. **g.** Raincloud plot of resistance and gauge factor measurements for 50 strain sensor units. The density plots, dot plots, and box plots show the distribution, median, and variability of resistance and gauge factor, respectively. **h.** Washability test under room temperature (RT) and 500 rpm stirring with a magnetic stir bar. Grey dashed lines show the maximum and minimum values measured during the experiment. **i.** The nasal breath signals during nocturnal turning collected from the device with (upper) and without (lower) strain isolation treatment.

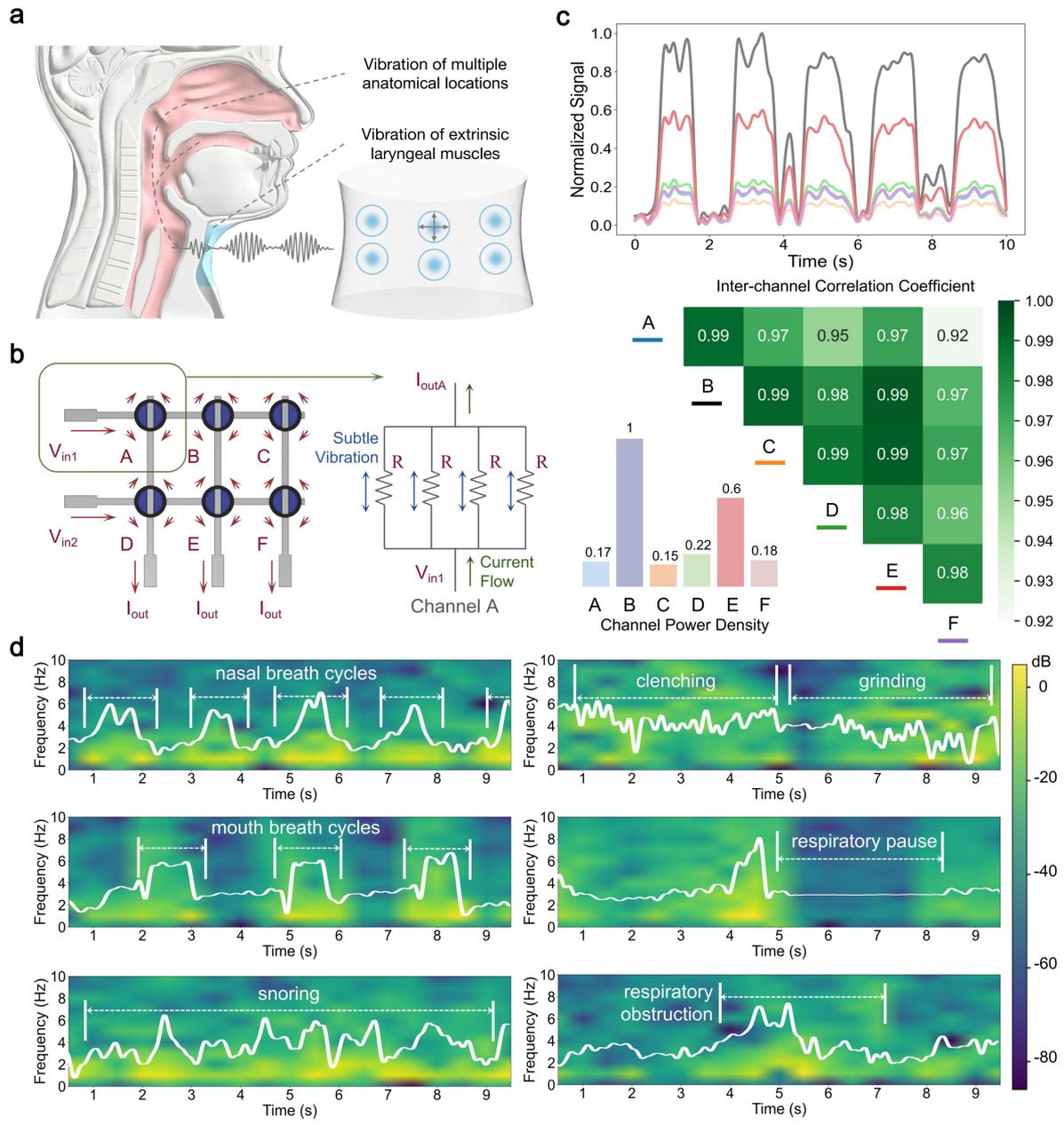

**Figure 3 | Performance of the strain sensor array in acquiring sleep conditions related signals. a.** Illustration of how sleep behavior generates extrinsic laryngeal vibrations. **b.** Equivalent circuit of the strain sensor array. **c.** Multichannel signal and channel correlation visualization. **d.** Time-frequency spectrogram of the channel signal with the strongest response.

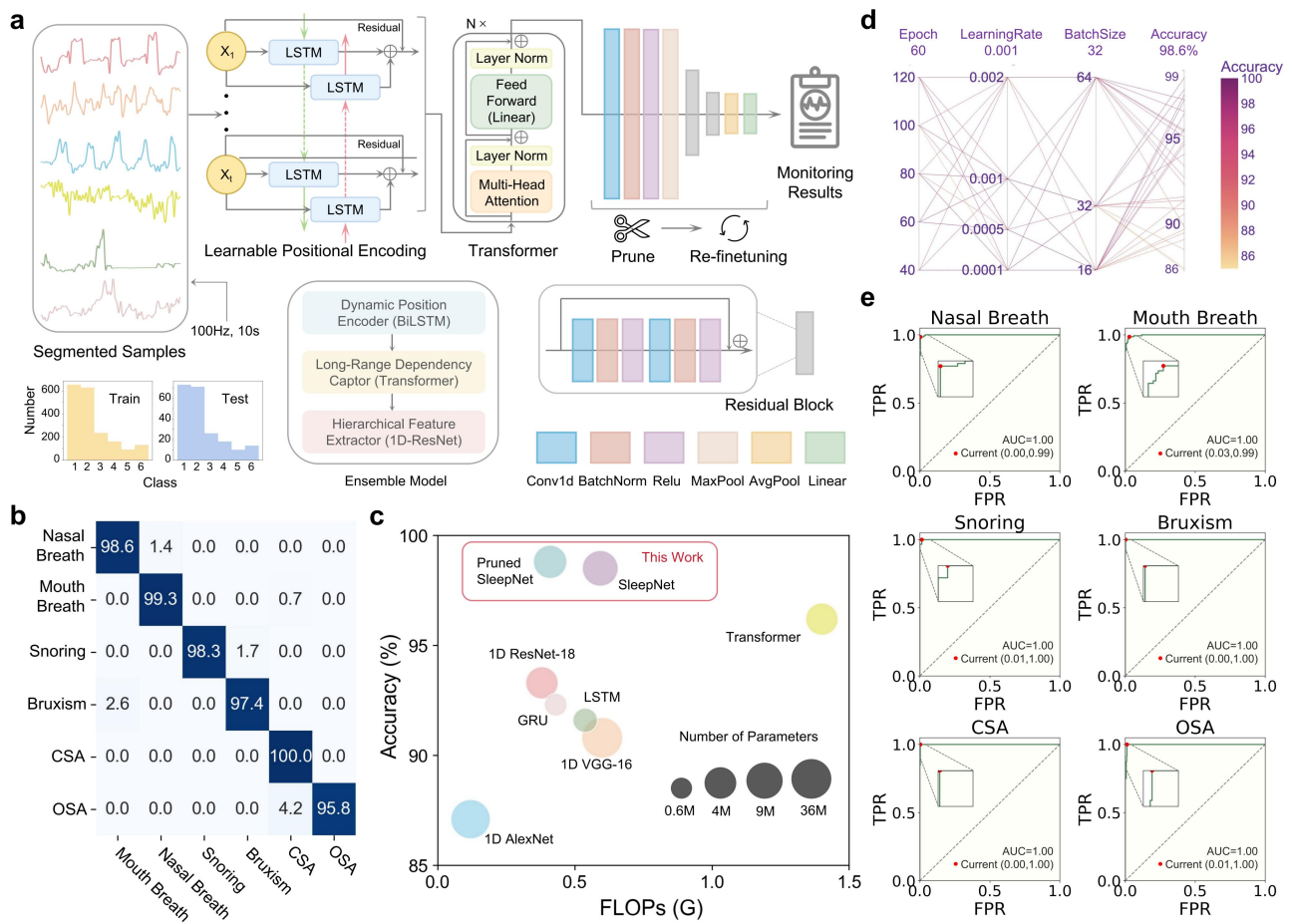

**Figure 4 | The sleep conditions recognition model. a.** Pipeline of the SleepNet for sleep pattern recognition. **b.** Confusion matrix showing the classification results for 6 sleep patterns. **c.** Comparison of model efficiency (measured in FLOPs), accuracy, and number of parameters with state-of-the-art neural network backbones. **d.** Process visualization of random hyperparameter optimization. **e.** Six receiver operating characteristic (ROC) curves and area under the curve (AUC) values of classification by the pruned SleepNet.

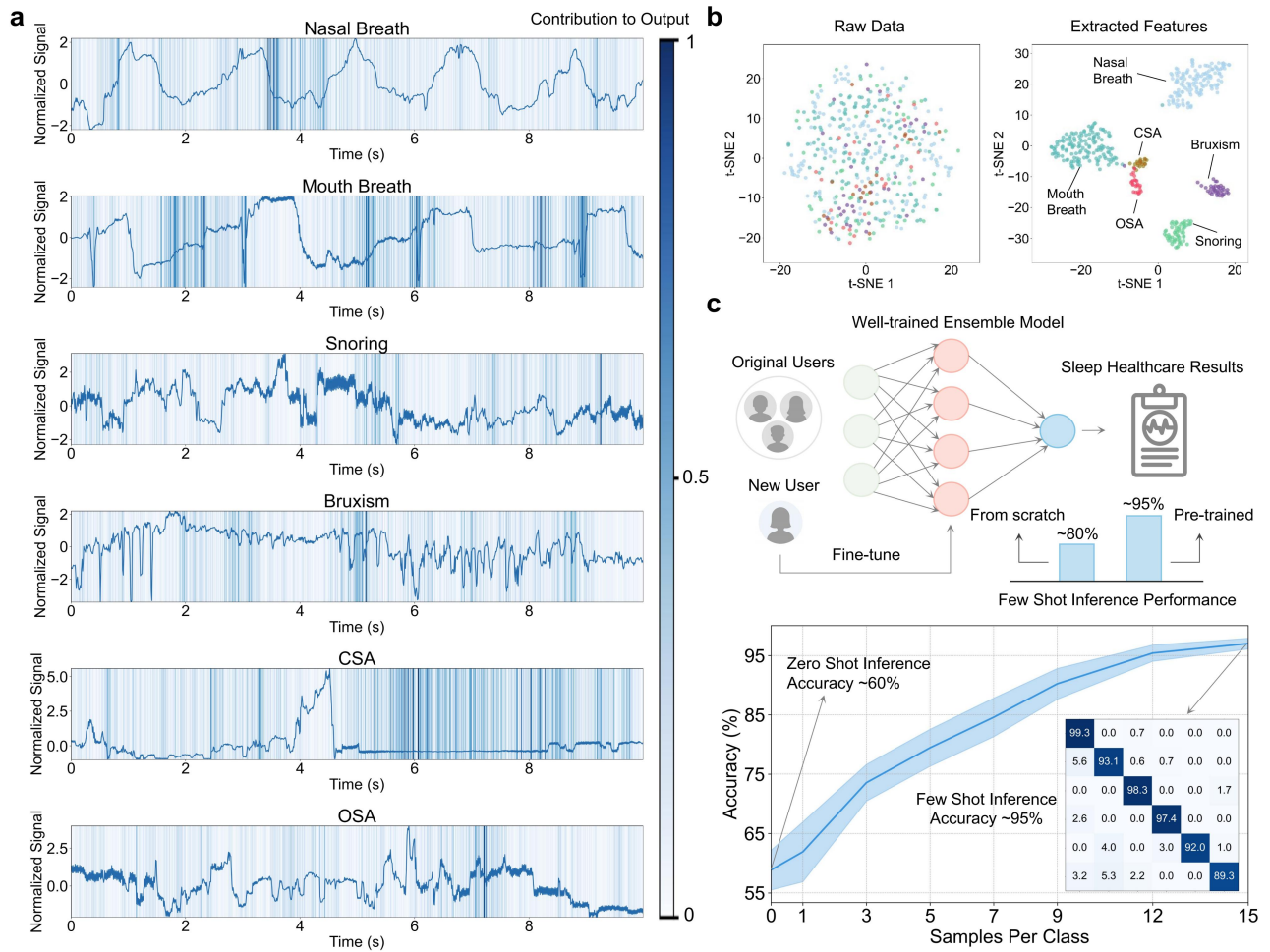

**Figure 5 | Explainability and generalizability analysis. a.** SmoothGrad displays the distribution of contributions that signals make to the model's classification output. **b.** T-distributed stochastic neighbor embedding (t-SNE) visualizations comparing the distribution of raw data to features extracted from the model. **c.** Flowchart and results depicting the model's generalization process.